\documentclass[fleqn,10pt,twocolumn]{wlscirep}
\usepackage[utf8]{inputenc}
\usepackage[T1]{fontenc}
\usepackage{hhline}
\usepackage{amsmath}
\usepackage{multirow}
\usepackage{subfig}
\usepackage[absolute]{textpos}
\title{Automated scoring of pre-REM sleep in mice with deep learning}

\author[1]{Niklas Grieger}
\author[2]{Justus T.~C.~Schwabedal}
\author[3]{Stefanie Wendel}
\author[3]{Yvonne Ritze}
\author[1,*]{Stephan Bialonski}
\affil[1]{Department of Medical Engineering and Technomathematics, FH Aachen University of Applied Sciences, Jülich, 52428, Germany}
\affil[2]{Independent researcher, Kottbusser Damm 35, Berlin, 10967, Germany}
\affil[3]{Department of Medical Psychology and Behavioral Neurobiology, University of Tübingen, Tübingen, 72076, Germany}
\affil[*]{bialonski@fh-aachen.de}

\begin{abstract}
Reliable automation of the labor-intensive manual task of scoring animal sleep can facilitate the analysis of long-term sleep studies. In recent years, deep-learning-based systems, which learn optimal features from the data, increased scoring accuracies for the classical sleep stages of Wake, REM, and Non-REM.
Meanwhile, it has been recognized that the statistics of transitional stages such as pre-REM, found between Non-REM and REM, may hold additional insight into the physiology of sleep and are now under vivid investigation.
We propose a classification system based on a simple neural network architecture that scores the classical stages as well as pre-REM sleep in mice.
When restricted to the classical stages, the optimized network showed state-of-the-art classification performance with an out-of-sample F1 score of $0.95$ in male C57BL/6J mice.
When unrestricted, the network showed lower F1 scores on pre-REM ($0.5$) compared to the classical stages. The result is comparable to previous attempts to score transitional stages in other species such as transition sleep in rats or N1 sleep in humans.
Nevertheless, we observed that the sequence of predictions including pre-REM typically transitioned from Non-REM to REM reflecting sleep dynamics observed by human scorers.
Our findings provide further evidence for the difficulty of scoring transitional sleep stages, likely because such stages of sleep are under-represented in typical data sets or show large inter-scorer variability.
We further provide our source code and an online platform to run predictions with our trained network.
\end{abstract}

\begin{document}

\flushbottom
\maketitle

\begin{textblock*}{21cm}(0cm,26.5cm)
  \centering
  This is a preprint of an article published in Scientific Reports 11, 12245 (2021).\\
  The final authenticated version is available online at: \href{https://doi.org/10.1038/s41598-021-91286-0}{10.1038/s41598-021-91286-0}
\end{textblock*}

\thispagestyle{empty}

\section{Introduction}
\label{sec:intro}

Sleep is one of the most fundamental and not yet well understood processes that can be observed in humans and most animals. Since the discovery of sleep-wake correlates in electroencephalographic (EEG) signals in the early 20th century, the sleep architecture (i.e., the temporal succession of different states of consciousness) has been under detailed investigation, and it is now known to affect critical processes such as autonomic function, mood, cognitive function, and memory consolidation\cite{Bashan2012, Rasch2013}. Essential to most sleep studies is the manual scoring of EEG recordings, which is known as a time-consuming, labor intensive task.  Automated scoring systems have been developed since the late 1960s\cite{Drane1969,Smith1969} to facilitate sleep scoring in humans\cite{Aboalayon2016,Faust2019,Fiorillo2019} and in animal models such as rodents\cite{Robert1999,Bastianini2015}.  While manual scoring remains the gold standard to evaluate sleep studies, such systems could potentially reduce problems such as inter-rater variability and analyst's fatigue introducing systematic errors into the derived sleep architecture.

In rodents, and in particular in mice, the development of systems to automatically score sleep focused on distinguishing between the main sleep stages Wake (wakefulness), REM (rapid eye movement), and NREM (non-rapid eye movement)\cite{Robert1999,Bastianini2015}. While approaches to automatically identify these stages have become increasingly successful, sleep research has uncovered finer structures including sub-stages in mice\cite{Katsageorgiou2018,Lacroix2018} and \emph{pre-REM} sleep\cite{Ruigt1989}, also known as \emph{intermediate stage}\cite{Glin1991,Gottesmann1992} or \emph{transition sleep}\cite{Mandile1996} in rats and mice\cite{Gottesmann1996,Amici2005}. Efforts to automatically identify such sleep patterns have been limited to rats, where studies have introduced systems to classify four to seven stages such as transition sleep\cite{Ruigt1989,Benington1994,Neckelmann1994,Gross2009,Wei2019}, different Wake states\cite{Ruigt1989,Kohtoh2008}, or different NREM stages\cite{Ruigt1989,Neckelmann1994,Wei2019}. In mice, however, published work to date has focused on scoring only the three main stages, with very few exceptions (different Wake states\cite{VanGelder1991}, cataplectic events\cite{Exarchos2020}).

Classical approaches towards automating sleep scoring in rodents are based on \emph{engineered features}\cite{Robert1999} which are often inspired by human sleep scorers who use visual representations of EEG signals for their scoring decisions. Power spectral densities (e.g., magnitudes and ratios of power in delta, theta, sigma bands \cite{Kohtoh2008,Rytkonen2011,Sunagawa2013,Bastianini2014,Caldart2020}), and EEG amplitudes (e.g., moments of EEG amplitude distributions\cite{Robert1999}) are among the most frequently used features. While systems based on carefully engineered features have seen progress in scoring performance, deep-learning-based systems that can create and use \emph{learned features} have repeatedly been demonstrated to achieve superior classification performance in related fields such as speech or image recognition\cite{LeCun2015,Goodfellow2016}. These successes indicate that learned features may be particularly effective and expressive representations of data with respect to the classification challenges at hand, as well as various EEG-based challenges including brain-computer interfaces, epileptic seizure detection, or sleep scoring\cite{Roy2019}.
\begin{table*}
\centering
\begin{tabular}{r|rrrrr}
 & training set & validation set & test set & all data\\\hline
 number of mice & 14 & 2 & 2 & 18\\
 recording days & 42 & 6 & 4 & 52\\
 number of segments & 366,720 & 52,140 & 35,420 & 454,280\\
 percentage & 80.7\,\% & 11.5\,\% & 7.8\,\% & 100\,\%
\end{tabular}
\caption{Characteristics of the dataset. Data from 18 mice were split into three disjunct sets (training, validation, and test set). To allow for robust assessment of generalization performance of our networks out-of-sample, data of each mouse was assigned to only one of these sets. Each recording was further split into segments of 10~seconds for which sleep stages (labels) were predicted.}
 \label{tab:1}
\end{table*}

While research into human sleep scoring has adopted ideas and methods from deep learning\cite{Fiorillo2019,Faust2019}, there are, to our knowledge, only few studies proposing deep-learning-based systems to improve animal sleep scoring, particularly in mice. Deep neural networks introduced in two studies\cite{Miladinovic2019,Barger2019} were inspired by image recognition systems and used preprocessed spectrograms of EEG and the electromyogram (EMG) as input ``images''. These systems yielded state-of-the-art classification performances but needed an additional hidden Markov model\cite{Miladinovic2019} that constrained the output of the system on physiologically plausible sleep state transitions or used mixture z-scoring\cite{Barger2019} which needs an initial sample of human-annotated data for each mouse. Four studies proposed deep neural networks that were trained end-to-end on raw input EEG and EMG signals. One of the earliest studies\cite{Yamabe2019} was based on 22 years of EEG and EMG signals from mice and reported an improvement of sleep scoring accuracy with respect to classical approaches. The system contained a bidirectional long short-term memory (LSTM) module in the classifier head to model long-range non-linear correlations\cite{Hochreiter1997}, and the training also included a retraining scheme. The second study\cite{Svetnik2020} reported a convolutional neural network to show better classification performance compared to a random forest trained on engineered features but was not based on human-annotated data (the study relied on software annotated data instead). Finally, the third study\cite{Exarchos2020} trained a convolutional neural network to predict sleep stages in narcoleptic and wild-type mice where cataplectic events were derived based on rules operating on the predictions (Wake, REM, or NREM) of the network. All discussed neural networks were trained to distinguish between the main sleep stages Wake, REM, and NREM only.

In this contribution, we introduce a neural network to automatically score the main stages Wake, NREM, and REM as well as pre-REM sleep and artifacts in mice with one EEG channel as input. To the best of our knowledge, this is the first contribution to study the capability of deep neural networks to score pre-REM in mice. We discussed challenges that arise from predicting pre-REM sleep including class imbalance and absence of consensus\cite{Bastianini2015} for scoring rodent sleep. When constraining our network to score the main stages only, our network showed state-of-the-art performance compared with previous work. We consider network architectures as proposed here and in other contributions\cite{Svetnik2020,Exarchos2020} to be particularly promising for promoting further development of sleep scoring systems. Finally, source code\cite{Grieger2020a} and an online platform\cite{Schwabedal2020a} are provided to allow for adapting and testing our trained network.

\section{Materials and methods}
\label{sec:methods}

\subsection{Data acquisition}

The dataset consists of polysomnographic recordings of 18 mice with a total recording duration of 52 days (corresponding to 454,301 windows (segments) of 10 seconds each, see table~\ref{tab:1}). Each mouse was recorded for 3 days, except for one mouse whose recording spanned 1 day because it unexpectedly passed away. The data were acquired during a study at the University of Tübingen which tested the influence of dietary variations on sleep.

\subsubsection{Animal nurturing and treatment}
All mice (age about 10 weeks, male, C57BL/6J) were kept at the animal facility FORS at the University of Tübingen, Germany. All applicable local and federal regulations of animal welfare in research were followed (Directive 86/609, 1986 European Community). Experiments were approved by the ethics committee of the Regional Council Tübingen, Germany, permit number MPV 3/15. Mice were housed and experiments were conducted at controlled temperature ($20 \pm 2$ C$^\circ$), humidity, and 12h light/dark cycle at all times (light on at 06:00 or 7:00 summer/winter time). Out of the 18 mice, 9 were fed with a different diet (sucrose solution instead of water) to test an experimental paradigm to be published elsewhere.
We did not consider the dietary groups as relevant for the present study and
did not attempt a statistical comparison thereof.  The mouse that passed away
unexpectedly was part of the altered-diet group.

\begin{figure*}
\centering
\includegraphics[width=\linewidth]{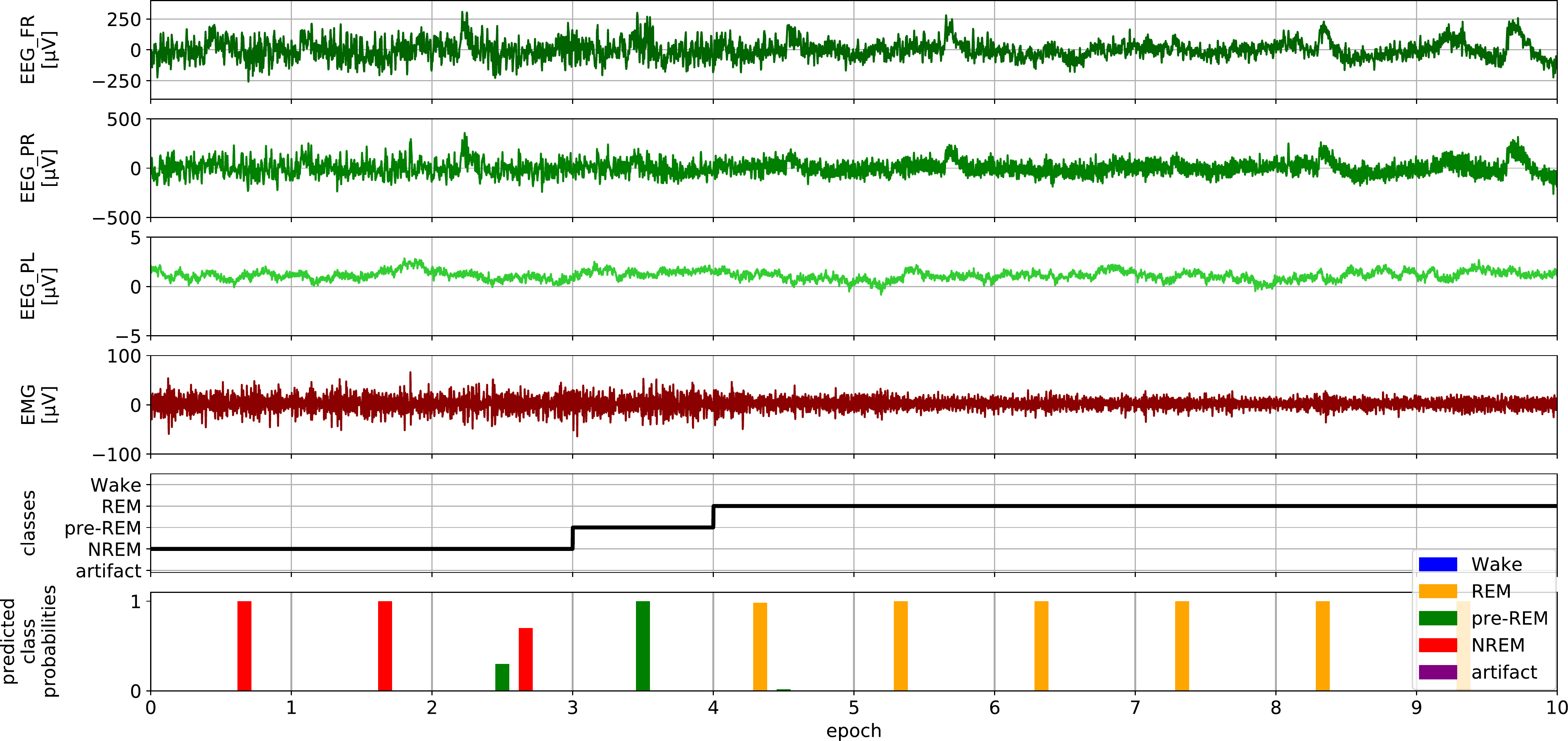}
\caption{Example EEG time series (panels 1--3: frontal, parietal right, parietal left) and EMG time series (panel 4) of a mouse measured by chronically implanted electrodes (data from the test set). Sleep scores are shown in panel 5, whereas the last panel shows score probabilities predicted by the network. Time series were split into segments of 10 seconds each (marked by vertical lines; see text). For each segment, the sleep score was determined by an expert, and score probabilities were predicted by our network.}
\label{fig:data}
\end{figure*}

\subsubsection{Polysomnographic recordings}
The mice were chronically implanted with a synchronous recording device of one EMG and four EEG electrodes seven days before the first recording. Four stainless steel screws were placed on top of the cortex using the following coordinates: frontal electrode (AP: +1.5 mm, L: +1.0 mm, relative to Bregma), two electrodes parietal left and right (AP: 2.0 mm, L: $\pm$2.5 mm), and the reference electrode occipital (AP: -1.0 mm, L: 0 mm, relative to Lambda) according to the atlas of Franklin and Paxinos\cite{Paxinos2019}. This setup was used to sample the electrical activity of the frontal, parietal-left, and parietal-right lobes. Two flexible stainless steel wires were inserted into the neck muscles to measure the EMG signal. During recordings, mice were plugged with a cable attached to a swiveling commutator, which was connected to the amplifier (Model 15A54, Grass Technologies, USA). The amplified (x1000) signals (EEG and EMG) were digitalized, filtered (EEG: 0.01--300 Hz; EMG: 30--300 Hz), and sampled at a rate of 992.06 Hz.
The commutator units were built in-house.  The software Spike2, version 6.07, was used for data acquisition and subsequent annotation.

\subsubsection{Manual sleep-stage classification}
Our expert (SW) manually scored sleep by visual inspection of EMG and EEG time series. The parietal right EEG was chosen for scoring decisions because it had the best signal-to-noise ratio across all mice. Recordings were divided into non-overlapping consecutive windows (segments) of 10 seconds. Three consecutive segments of the EMG and the parietal right EEG signal were visually presented to the expert who scored the middle segment. The expert was not required to score the recordings in sequential order but could jump to other parts of the recordings at all times.
Epochs containing features of multiple sleep stages were assigned the stage
that made up the majority of the epoch.

Segments were scored as stage ``Wake'', ``NREM'', ``pre-REM'', or ``REM'' sleep (see figure~\ref{fig:data} and table~\ref{tab:datasets}). Artifact contaminated segments were scored as ``artifact''. The Wake state is characterized by elevated activity in the EMG signals when compared with NREM, pre-REM, and REM sleep. NREM sleep is characterized by slow waves between 0.5--4\,Hz (delta band) while REM sleep is associated with rhythmic activity between 7--8\,Hz (theta band). The pre-REM stage is characterized by successive short bouts ($< 10$\,s) of high amplitudes similar to spindles and the appearance of a theta rhythm~\cite{Glin1991,Amici2005}. Wake was the most frequent stage in our dataset ($55.08\,\%$, see table~\ref{tab:datasets}), followed by NREM ($37.7\,\%$) and REM sleep ($5.08\,\%$). The intermediate stage pre-REM occurred only rarely ($1.88\,\%$) and only $0.26\,\%$ of all segments were contaminated with artifacts.

\begin{table*}
\centering
\tabcolsep=0.11cm
\begin{tabular}{r|rrrrrr}
 label & Wake & REM & NREM & pre-REM & artifact \\\hhline{=|======}
 \emph{all data} & & & & & &\\
 \# segments & 250,206 &  23,065 & 171,270  & 8,536  & 1,203  \\
 perc. &  55.08\,\% &  5.08\,\% &  37.70\,\% &  1.88\,\% &  0.26\,\%\\\hhline{=|======}
 \emph{train. set} & & & & & &\\
  \# segments & 203,747 & 18,606 & 136,586 & 7,181 & 600\\
 perc. & 55.56\,\% & 5.07\,\% & 37.25\,\% & 1.96\,\% & 0.16\,\% \\\hhline{=|======}
 \emph{reb. train. set 1} & & & & & &\\
 \# segments & 110,017 & 66,010 & 88,013 & 51,341 & 51,341  \\
 perc. & 30\,\% & 18\,\% & 24\,\% & 14\,\% & 14\,\% \\\hhline{=|======}
  \emph{reb. train. set 2} & & & & & &\\
 \# segments & 150,110 & 91,531 & 124,481 & - & - \\
 perc. & 41\,\% & 25\,\% & 34\,\% & - & -\\\hhline{=|======}
 \emph{validation set} & & & & & &\\
  \# segments & 28,382 & 2,544 & 19,952 & 752 & 510\\
 perc. & 54.43\,\% & 4.88\,\% & 38.27\,\% & 1.44\,\% & 0.98\,\% \\\hhline{=|======}
 \emph{test set} & & & & & &\\
  \# segments & 18,077 & 1,915 & 14,732 & 603 & 93\\
 perc. & 51.04\,\% & 5.41\,\% & 41.59\,\% & 1.70\,\% & 0.26\,\%
\end{tabular}
 \caption{Statistics of labeled segments of the whole dataset (``\emph{all data}''), of the training set (``\emph{train. set}''), the rebalanced training sets (``\emph{reb. train. set 1} and \emph{2}''), the validation set (``\emph{validation set}''), and the test set (``\emph{test set}''). NREM: Non REM sleep, pre-REM: Pre REM sleep, Wake: Wakefulness. perc: percentage,  \#\,segments: number of segments.}
 \label{tab:datasets}
\end{table*}

\subsection{Dataset preparation for machine learning}

\subsubsection{Train-, validation-, and test-set splits}
\label{sec:splits}

The scored time series were prepared for training and evaluation of the neural network for automated sleep scoring as follows: To mimic the manual scoring process and make our automated scoring system more flexible with respect to different experimental setups (in which only one EEG derivation may be available), we chose to base the training of the neural network on the parietal right EEG only. The parietal right EEG time series were low-pass filtered using a 4-th order Butterworth low-pass filter (critical frequency: 25.6 Hz) in order to prevent aliasing effects due to subsequent downsampling. The low-pass filter was applied to each time series with a forward- and backward pass to avoid any phase shifts. Subsequently, the low-pass filtered time series were downsampled to 64 Hz by linear interpolation. 

The recordings were assigned to a training, a validation, and a test set (see tables~\ref{tab:1} and~\ref{tab:datasets}).
Subjects were distributed such that each set consisted of recordings from mice of both dietary groups. Validation and test set each consisted of recordings from two mice one of which was from the altered-diet group.  The
mouse that passed away early was assigned to the test set so that later evaluation would
allow us to test the trained network on the most novel subject in our dataset.

The network parameters were trained on the training set, whereas the validation set allowed us to optimize hyperparameters of our network, of the training process and of regularization as well as data augmentation strategies. Finally, the test set was used to assess the performance of our network to predict sleep stages out-of-sample (i.e., on data from mice that were not available during training nor validation). We stress the importance of out-of-sample testing since it mimics the situation in sleep laboratories where a trained network predicts sleep stages on data from mice that are unknown to the network. To simulate this situation, we made sure that all data from each mouse was assigned to one and only one of the three sets (see table~\ref{tab:1}).

We note that, when the network was trained to predict only the standard set of target stages (Wake, NREM, REM), the training, validation, and test set were created by reinterpreting pre-REM stages as NREM, while segments scored as artifact were removed.

\subsubsection{Class rebalancing}

The sleep stages identified by the expert occur with widely differing frequencies (e.g., Wake: $55.08\,\%$ vs pre-REM sleep: $1.88\,\%$, see first row of table~\ref{tab:datasets}). Such differing frequencies of classes (\emph{class imbalance}) can pose non-trivial challenges when creating data-driven systems due to the tendency of systems to over-classify majority classes and misclassify infrequent classes~\cite{Haixiang2017,Johnson2019}. To address this issue, we created rebalanced training sets that differed in their class frequencies (see rows 3 and 4 of table~\ref{tab:datasets}) while maintaining the total number of segments in the set. The rebalanced training sets were created before each training epoch by random sampling with replacement from the respective classes of the original training set so that predefined class frequencies were reached. In the rebalanced training set \#1, class frequencies were heuristically chosen such that infrequent classes were oversampled and frequent classes undersampled while class frequencies still reflected some of the class imbalance of the original training set. The rebalanced training set \#2 was created to investigate the performance of our network to distinguish between the three main stages (Wake, NREM, REM) only.

\subsubsection{Data augmentation}
\label{sec:data-augmentation}

Creating additional synthetic training data is known as \emph{data augmentation} and has repeatedly been demonstrated to reduce overfitting and to tackle class imbalance\cite{Fawaz2019,Wen2020,Iwana2020}. Moreover, if prior knowledge about the data cannot be easily incorporated into the design of a network architecture, the knowledge can often be utilized to create synthetic training data. This way prior knowledge can indirectly be induced into the network during training. A prime example is the creation of rotation invariance in image recognition systems by supplying rotated images (derived from the original data) during the training process\cite{Goodfellow2016}.

We expect variability in our data that does not reflect sleep dynamics but can be introduced by the data acquisition, the scoring process, or unrelated physiological processes. Our data augmentation strategies are aimed at inducing invariance in our network with respect to this variability.

(i) Amplitude variability. Different acquisition systems can use different signal amplification settings, thereby leading to a variability in the scaling of signal amplitudes. Let $\vec{s}$ denote the part of the time series that is presented to the input layer of the neural network during training. Let $s_t$ denote the sample of this time series at time step $t\in[1,T]$, where $T$ denotes the length of the time series. During training, instead of $\vec{s}$ we presented the network with the augmented time series
\begin{equation}
 \vec{s}^* = \vec{s}(1+a_a \cdot u) \qquad \qquad u \in U[-1, 1),
\end{equation}
where $u$ is a random variable drawn from the continuous uniform distribution in the interval $[-1, 1)$, and where $a_a \in [0, 1]$ is the strength of this type of augmentation. $u$ is newly drawn for each time series entering the network.

(ii) Frequency variability. We expect that the frequency content of EEG signals varies to a certain extent independently of the actual sleep stage, due to different noise sources (e.g., caused by movements of mice or by the measurement device itself). We mimicked such a variability in frequency content by using \emph{window warping}\cite{LeGuennec2016} as an augmentation technique in which a time series is stretched or contracted. Let $T$ denote the length of the time series that enters the network. To create an augmented time series, in a first step, we determined a temporary length
\begin{equation}
 T^* = \lfloor T(1+a_f\cdot u) \rfloor \qquad u \in U[-1,1)
\end{equation}
of a temporary time series, where $a_f \in [0, 1]$ denotes the strength of the stretching or contracting. The central sample point of the temporary time series is identical to the one of the original time series. In a second step, the temporary time series was resampled by linear interpolation such that its new length equals $T$. If $T^* > T$, the temporary time series includes sample points prior to and after the original time series. In this case, the resampling operation contracts the time domain and thus frequencies are shifted towards higher values. If $T^*<T$, the resampling operation stretches the time domain and frequencies are shifted towards lower values.

(iii) EEG montage variability (sign flip). Since each EEG signal represents a difference between the electrical potentials at two (or more) measurement sites, the sign of the amplitudes of an EEG signal depends on the chosen \emph{montage} (i.e., the configuration of electrical potential differences between electrodes that is used to represent EEG signals). A change in montages can lead to a flipped sign of the amplitudes of the signal which, however, still reflects the same brain dynamics. The augmented time series $\vec{s}^*$ reads
\begin{equation}
 \vec{s}^* = \begin{cases}-\vec{s} & \text{if }u<a_s\\ \vec{s} & \text{else}\end{cases}\qquad u \in U[0, 1),
\end{equation}
where $a_s\in[0,1]$ is the probability of flipping the sign of the time series amplitudes.

(iv) Time shift variability. Sleep scoring associates a sleep stage to each of the consecutive EEG segments.  Thus, the segment length introduces an artificial timescale on which sleep is characterized. Actual sleep stages do not necessarily change at segment boundaries but can change within a segment. To account for this variability, we created augmented data by shifting the time series segments with respect to the sleep scores by a random time shift $\Delta t$,
\begin{equation}
\Delta t = \lfloor T \cdot a_t\cdot u \rfloor \qquad u\in U[-1,1],
\end{equation}
where $a_t \in [0, 1]$ controls the extent of the shift and thus the strength of this type of augmentation. With $\vec{s}=(s_1, \ldots, s_T)^\text{T}$ denoting the original time series, the augmented time series reads
\begin{equation}
 \vec{s}^* = (s_{1+\Delta t}, \ldots s_{T+\Delta t})^\text{T}.
\end{equation}

Time series were augmented in the order (i), (iii), (ii), (iv) as described above using newly drawn random variables every time the time series entered the neural network for forward propagation. The strength of the augmentations were controlled by the parameters $(a_a, a_f, a_s, a_t)$.

\begin{figure*}[h]
\centering
\includegraphics[width=0.9\linewidth]{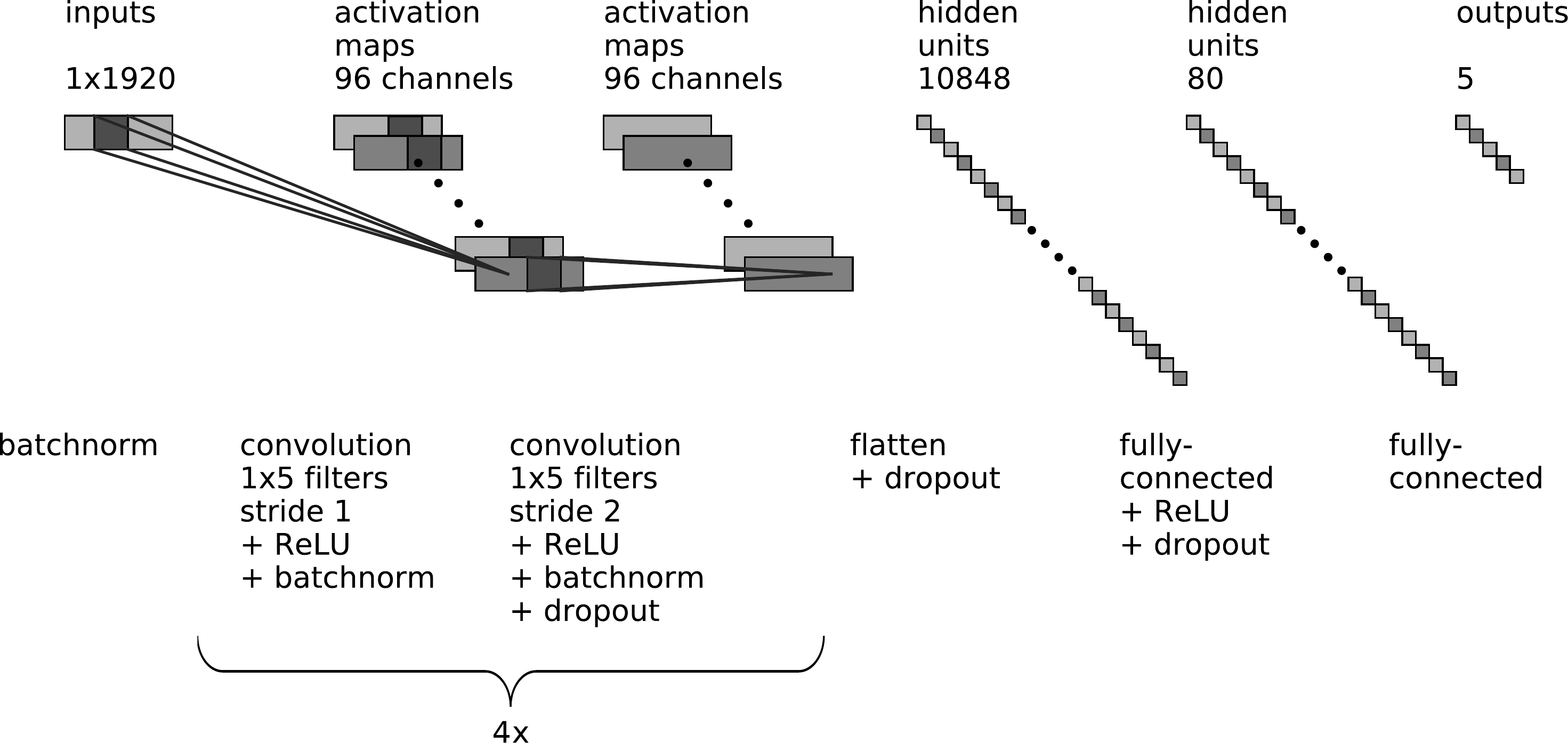}
\caption{Network architecture. The \emph{feature extractor} consists of 8 convolutional layers, and the output of the last convolutional layer is flattened (Flatten). EEG time series data is batch normalized (Batchnorm) before the first convolutional layer is applied. The \emph{classifier} consists of two fully connected layers. A softmax layer translates the activations of the previous layer into class probabilities.}
\label{fig:network}
\end{figure*}

\subsection{Machine learning}

\subsubsection{Network architecture}

Mimicking the manual scoring process, the input of the neural network consists of three consecutive segments of the parietal right EEG time series, while the outputs of the network are the probabilities of the middle segment to belong to Wake, NREM, pre-REM, REM sleep, or to contain artifacts (class artifact). We used the score with the highest probability as the predicted score by our network.

The network architecture, which draws upon experience gained in previous work by some of the authors\cite{Schwabedal2018b}, consists of a feature extractor and a classifier (see figure~\ref{fig:network}). Before entering the feature extractor, the time series is batch normalized (``Batchnorm'', see figure~\ref{fig:network}). The feature extractor consists of 8 convolutional layers, each of which is composed of 96 kernels of size $d\times 1\times5$, where $d$ denotes the depth of the output volume of the previous layer. Since one time series enters the network, $d=1$ for the first convolutional layer, while $d=96$ for all other convolutional layers. To downsample the time series information along the feature extractor, every other convolutional layer uses a stride of 2 with the other layers having a stride of 1. The resulting convolved signals of each layer are nonlinearly transformed by rectified linear units (ReLUs) and subsequently batch normalized (Batchnorm). Furthermore, we apply Dropout regularization\cite{Srivastava2014} to every other convolutional layer. Finally, the output (also called features) of the feature extractor is concatenated into a vector (``Flatten'', see figure~\ref{fig:network}), and Dropout regularization is applied before these features enter the classifier.

The classifier is composed of 2 fully connected (FC) layers with $113 \cdot 96 = 10848$ and $80$ neurons, respectively. The first FC layer is equipped with rectified linear units (ReLUs) as nonlinearity and Dropout regularization, while the second FC layer uses a softmax activation function to determine the score probabilities $\vec{p}$ which represent the activations of the neurons of the output layer. The output layer consisted of 3 output neurons when the network was trained to distinguish between major sleep stages (Wake, REM, NREM). When the network was trained to also score pre-REM sleep and segments containing artifacts, the output layer consisted of 5 neurons.

\subsubsection{Training}

\emph{Loss function.} The training objective is implemented by the loss function $\mathcal{L}$ that becomes small when the classifier assigns a large probability to the correct class. Let $ \vec{x}_i$ denote a time series segment $i$ and let $k$ be the index of the middle segment of the input time series for which the network predicts the score probability vector $\vec{p}_{k}$. With $c_k\in\{1, \ldots, 5\}$ we encoded the correct score for segment $k$ as determined by our human expert. Thus, $p_{k, c_k}$ is the predicted probability of the middle time series segment to belong to the sleep stage that was determined by our human expert. The loss is the negative log-likelihood function with an L2 regularization penalty term $L_2$,
\begin{equation}\label{eq:loss_function}
 \mathcal{L} = -\frac{1}{N_b}\sum\limits_{k=1}^{N_b}\log{(p_{k,c_k})} + \underbrace{ \frac{\lambda}{2N_b} \sum\limits_{w} w^2}_{L2 \text{ penalty}},
\end{equation}
where $N_b$ is the size of the minibatch, $\lambda\geq0$ is the L2 regularization strength, and where the sum of the L2 penalty is over all the weights $w$ in the network (except the bias weights).

\emph{Gradient descent.} Since the neural network is differentiable, the loss function can be minimized by minibatch gradient descent. The network was trained by Adam~\cite{Kingma2015}, a variant of stochastic gradient descent which uses exponential moving averaging of the gradient (equation \eqref{eq:gradient}) and the squared gradient (equation \eqref{eq:sqgradient}) to allow for adaptive learning rates. Let $w_t$ denote a weight (i.e., a free parameter) of the network and let $g_t=\frac{\partial \mathcal{L}}{\partial w}$ denote the gradient of the loss function with respect to the weight obtained for the minibatch at step $t$ of the training. The weight is updated by
\begin{equation}\label{eq:update_rule}
 w_{t+1} = w_t - \eta \cdot\frac{  \hat{m_t}}{\sqrt{\hat{v_t}} + \epsilon}
\end{equation}
where $\eta$ denotes the learning rate, and $\hat{m_t}$ and $\hat{v_t}$ are defined as
\begin{equation}\label{eq:gradient}
 \hat{m_t} = \frac{m_t}{1-\beta_1^t} \text{ with } m_t = (1-\beta_1)g_t + \beta_1 m_{t-1}
\end{equation}
and
\begin{equation}\label{eq:sqgradient}
 \hat{v_t} = \frac{v_t}{1-\beta_2^t}  \text{ with } v_t = (1-\beta_2)g_t^2 + \beta_2 v_{t-1},
\end{equation}
respectively. We used standard values for the exponential decay rates $\beta_1=0.9$ and $\beta_2=0.999$ and set $\epsilon=10^{-8}$ to prevent division by zero in equation~\eqref{eq:update_rule}. Furthermore, to counteract the exploding gradient problem, we employed a common gradient clipping strategy\cite{Pascanu2013} that ensured by rescaling that the norm of each gradient vector does not exceed a threshold $\theta=0.1$.

\emph{Learning rate protocol.} To speed up the learning process, we used large minibatches ($N_b=256$) and adjusted the learning rate $\eta$ according to the following protocol: The learning rate was linearly increased\cite{Goyal2017} from $10^{-7}N_b$ to $10^{-6}N_b$ over the course of the first 12 \emph{training epochs} (warm-up period). This type of linear scaling of the learning rate with minibatch size in combination with a warm-up period was observed to facilitate the use of large minibatch sizes, thereby shortening training times\cite{Goyal2017}. A single training epoch was completed when all minibatches of the training set were used once for the gradient updates. The warm-up period was followed by a cool-down period during which the learning rate was exponentially decreased, $\eta_i = 10^{-6}N_b\cdot e^{-\alpha (i-12)}$, where $i>12$ denotes the training epoch, and where the exponential rate was set to $\alpha=0.06$. The cool-down period lasted for the remaining training process.

\emph{Regularization.} We used four mechanisms to prevent our network from overfitting the training data. (i) \emph{L2 regularization}, also known as \emph{weight decay} (see second term in equation~\eqref{eq:loss_function}), favors neural networks whose weights (parameters) take on low values, thereby effectively restricting the model space of networks where the loss function obtains low values\cite{Montavon2012}. We set the L2 regularization strength to $\lambda = 10^{-4}$. (ii) We used \emph{Dropout regularization} with a dropout probability of $p_{dropout}=0.2$ in the feature extractor as well as in the classifier of the network (see figure~\ref{fig:network}). Dropout regularization has been reported to prevent complex co-adaptation of neurons and thus drive them to create features on their own\cite{Srivastava2014}.
(iii) Limiting the search in parameter space during training (called \emph{early stopping}) has a regularizing effect since it prevents the network from overfitting the training data\cite{Montavon2012}: After each training epoch, the network is evaluated on the validation set by calculating the F1 score (see section~\ref{sec:evaluation_measures}). If the F1 score does not improve over 5 consecutive training epochs, the training is stopped. The training is carried out for a minimum of 12 and a maximum of 50 epochs. At the end of the training, the network parameters are returned at that point of the training for which it achieved the best F1 score on the validation set. (iv)  Using prior knowledge to create additional training instances is called \emph{data augmentation}. Data augmentation is known to decrease the generalization error of the trained network and thus can be interpreted as a regularizing mechanism. We detail this approach in the section on data augmentation.

We note that we chose regularization parameters for all mechanisms after extensive hyperparameter exploration on the validation set.

\subsubsection{Evaluation measures}
\label{sec:evaluation_measures}
To investigate and assess the accuracy of our classifiers to score sleep, we used the following evaluation measures.

\emph{Confusion matrix.} We use confusion matrices to summarize the predictive performance of our networks.
The matrix elements describe the number of one of the actual labels (vertical dimension) that have been assigned one of
the target labels (horizontal dimension) by the network.
Both the actual label frequency (vertical axis) and the predicted label frequency (horizontal
axis) can be used to normalize the matrix elements (see figure~\ref{fig:confusion_matrices} for an
example)\cite{Ting2017}.
When normalized by the actual label frequency, matrix elements
show the percentage of segments of a true class being labeled by the network as a predicted class. Each diagonal element
corresponds to the network's recall of the respective class. Likewise, when normalized by the predicted label frequency,
matrix elements show the percentage of all segments of a class predicted by a network to actually belong to the true class.
Each diagonal element corresponds to the network's precision of the respective class.
\begin{figure*}[h]
 \centering
 \includegraphics[width=0.9\linewidth]{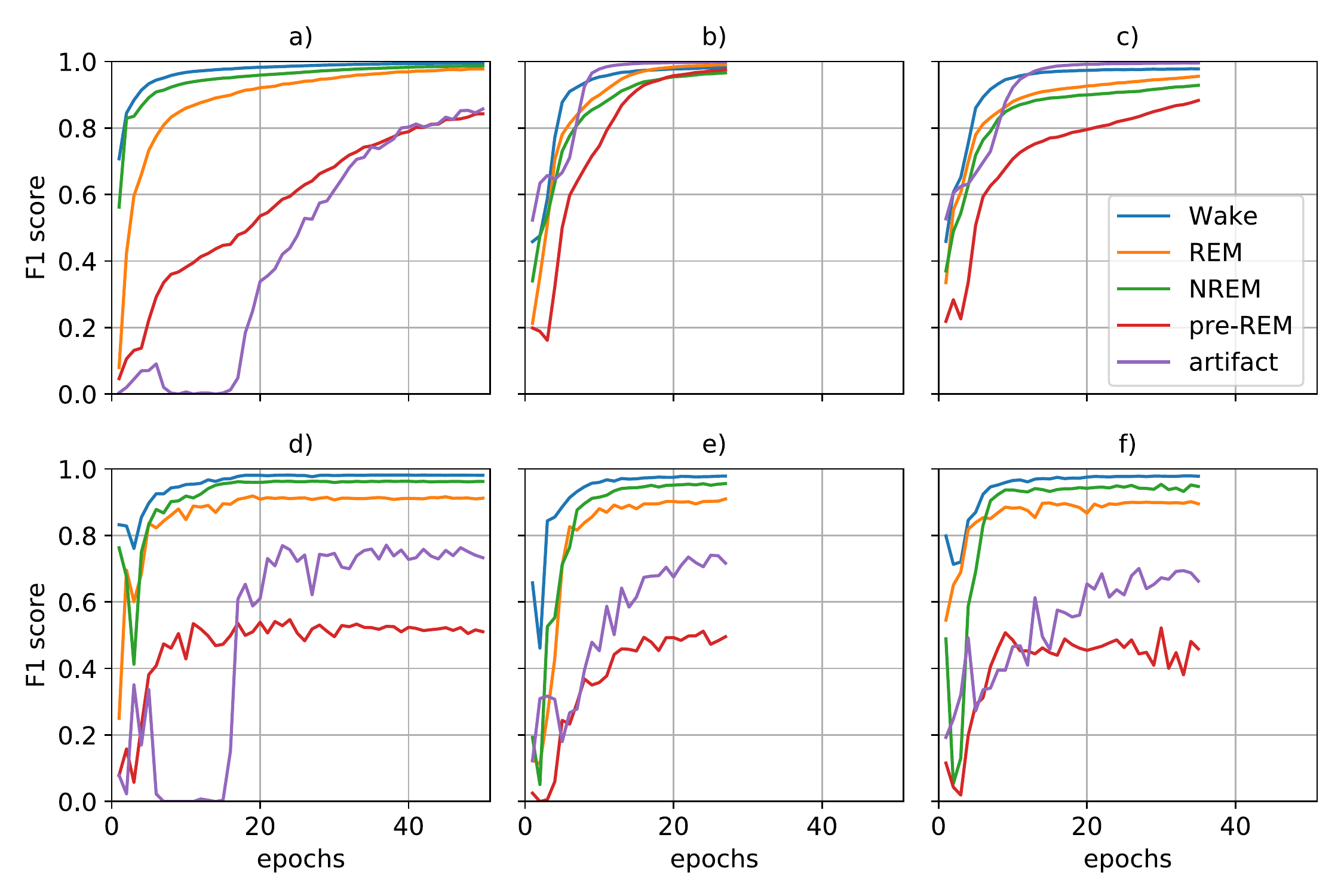}
 \caption{Evolution of F1 scores during training determined on the training set (top row) and validation set (bottom row). The evolution of F1 scores changed depending on whether the network was trained on unbalanced training data (panels a and d), rebalanced training data \#1 (panels b and e), or rebalanced training data \#1 with time shift augmentation (panels c and f).}
 \label{fig:f1vis}
\end{figure*}

\emph{F1 score.} In a binary classification setting (e.g., Wake vs Non Wake; pre-REM vs Non pre-REM) the F1 score, also
called F-measure, is defined as the harmonic mean of precision ($p_{precision}$) and
recall ($p_{recall}$)\cite{Tharwat2018}. Estimates of these values ($\hat{p}_{precision}$, $\hat{p}_{recall}$) are extracted from the normalized confusion
matrices:
\begin{equation} \label{eq:f1score}
    F1 = 2 \frac{\hat{p}_{precision} \cdot \hat{p}_{recall}}{\hat{p}_{precision} + \hat{p}_{recall}} = 2 \left( \frac{1}{\hat{p}_{precision}} + \frac{1}{\hat{p}_{recall}} \right)^{-1}
\end{equation}
The score takes on values between 0 and 1, with high values indicating better classification performances.
Although this metric is not invariant to changes in the data distribution, it is commonly used to compare different algorithms
against a consistent dataset\cite{Tharwat2018}. Note that rebalancing of the training set affects the training score
in unpredictable ways; however, we compare our networks based on the F1 scores calculated on the validation set which is
not rebalanced.
We expanded the definition given in equation~\eqref{eq:f1score} for our multiple class classification problem by
calculating $\overline{F1}$, the average of the F1 scores of each sleep stage.

\emph{Markov transition matrix.} Markov transition matrices summarize the empirical transition probabilities of a sequence
(e.g., a sequence of sleep stages scored by a human expert or a model) interpreted as a Markov process. They are two-dimensional
matrices whose elements describe the number of transitions from one label to another in either absolute numbers or,
after normalization, percentages (see figure~\ref{fig:tm} for an example). We created Markov transition matrices for
the actual labels and the labels predicted by our networks to measure how well the networks perceived the structure
of the sleep architecture.

\section{Results}

We investigated how accurate our network predicts sleep stages under various
training configurations.
We extensively modified the network capacity,
the training schedule, class rebalancing weights, and data augmentation.
Our experiments were conducted on two sets of target stages;
a standard set containing Wake, NREM, and REM stages,
and an extended set also including the stages pre-REM and artifact.
Our approach was to optimize hyperparameters on the extended set of sleep
stages.  Afterwards, we translated these parameters to the standard set to
investigate if the predictions reproduced the known state-of-the-art
performance.

In preliminary experiments, we also considered methods of cross-validation in
order to reduce the variance of predictions on the validation set.  We
discovered, however, that the reduction in variance did not yield additional
information that would influence our parameter search.  We therefore abandoned
this compute-intensive approach in favor of exploring a broader set of
parameters.
Note, that we did not cross-validate across samples in the test
set as this would have invalidated the generalization of our final results.

\subsection*{Influence of class rebalancing}

To analyze class rebalancing, we compared the epoch-wise improvements in F1 score
on the unbalanced and rebalanced dataset (see figure~\ref{fig:f1vis} panels (a,d) and (b,e) respectively). When trained on the unbalanced training set, F1
scores increased rapidly for the majority classes during the first 10 training
epochs, while the F1 scores of the pre-REM and artifact classes only slowly increased.
A similar behavior was observed for the F1 scores on the validation set.

Meanwhile, the F1 scores of the pre-REM and artifact classes obtained when training on the rebalanced training set increased more quickly
and reached higher optimal values. The optimal F1 scores on the validation set were comparable to the F1 scores obtained
on the validation set by the network trained on the unbalanced training set.

While the optimal F1 scores on the validation set were comparable between trainings on the rebalanced and unbalanced
training sets, we observed that training on the unbalanced dataset led to
\textit{unstable} results: minority-class F1 scores were sometimes not
recognized in lapses of several epochs. Such an event is visible for the F1
score of the artifact class in figure~\ref{fig:f1vis}(a,d).
Because of this instability, we continued to optimize the training configuration
using only rebalanced training sets.
\begin{table*}[h]
  \begin{center}
    \begin{tabular}{c|c|c|c|c}
	    training configuration & dataset & F1 score & F1 score & F1 score\\
        & & (mean) & pre-REM & artifact\\
        \hline
        \multirow{2}{*}{\shortstack{no regularization,\\no data augmentation}}
        & train & 0.99 & 0.99 & 1.00\\
        & valid & 0.76 & 0.44 & 0.56\\
        \hline
        \multirow{3}{*}{\shortstack{best regularization,\\no data augmentation,\\\textit{Best Network}}}
        & train & 0.98 & 0.96 & 1.00\\
        & valid & 0.81 & 0.50 & 0.74\\
        & \textbf{test} & \textbf{0.78} & \textbf{0.48} & \textbf{0.59}\\
        \hline
        \multirow{2}{*}{\shortstack{amplitude\\ $a_a=0.5$}} & train & 0.98 & 0.96 & 1.00\\
        & valid & 0.81 & 0.49 & 0.74\\
        \hline
        \multirow{2}{*}{\shortstack{frequency\\ $a_f=0.05$}} & train & 0.97 & 0.95 & 1.00\\
        & valid & 0.81 & 0.49 & 0.72\\
        \hline
        \multirow{2}{*}{\shortstack{sign flip\\ $a_s=0.5$}} & train & 0.97 & 0.94 & 1.00\\
        & valid & 0.80 & 0.49 & 0.69\\
        \hline
        \multirow{2}{*}{\shortstack{time shift\\ $a_t=0.06$}} & train & 0.94 & 0.86 & 1.00\\
        & valid & 0.81 & 0.52 & 0.67\\
        \hline
        \multirow{2}{*}{\shortstack{all data augmentations\\combined}} & train & 0.92 & 0.79 & 0.99\\
        & valid & 0.79 & 0.50 & 0.64\\
    \end{tabular}
    \caption{F1 scores for the training, validation, and test set at various training configurations. First, the network was trained without L2 regularization and without dropout (row 1). Next, the network was trained with both regularizations (row 2). Last, the network was trained using different types of data augmentation (rows 3--6; the parameters in column 1 resulted in the best validation F1 scores among the explored parameters), or with all data augmentations combined (row 7). F1 scores averaged over all classes (column 3) characterize how well the networks can predict sleep stages on all five classes. F1 scores in columns 4 and 5 present the capabilities of the networks to predict the two minority classes pre-REM and artifact, respectively. Only the network with the best validation F1 scores (best network) was evaluated on the test set.}
    \label{tab:f1ScoresDataAug}
  \end{center}
\end{table*}

\subsection*{Classification on the extended set of sleep stages}

We studied the classification performance under different training
configurations when targeting pre-REM and artifact in addition to the standard
classes Wake, Non-REM, and REM (see~table~\ref{tab:f1ScoresDataAug}). Without regularization and augmentation, we
observed a mean F1 score of $\overline{F1}=0.99$ on the training set, and $\overline{F1}=0.76$ on the
validation set. We optimized this F1 score by changing the dropout probability
$p_{dropout}\in\left[0.1,\mathbf{0.2},0.3,0.4,0.5,0.6,0.7,0.8\right]$ and
L2 regularization strength $\lambda\in\left[10^{-1},10^{-2},10^{-3},\mathbf{10^{-4}}\right]$
and achieved an improved mean F1 score of $\overline{F1}=0.81$ on the validation set (optimal values of hyperparameters
indicated in bold).
To increase the variability of the training data and reduce the possibility of overfitting the minority classes, we
continued by applying four data augmentation strategies (see section~\ref{sec:data-augmentation}) with parameters
in the following ranges:
\begin{itemize}
	\item Amplitude augmentation: $a_a\in[0.5,0.75,1.0]$
	\item Frequency augmentation: $a_f\in[0.05,0.1,0.2,0.5]$
	\item Sign flip augmentation: $a_s\in[0.25,0.5,0.75]$
	\item Time shift augmentation:\\
    $a_t\in[0.01,0.02,0.03,0.06,0.1]$
\end{itemize}

Results of the trainings with data augmentation are reported in table~\ref{tab:f1ScoresDataAug}. We observed a slight
decrease of mean F1 scores on the training set with augmentation (e.g., $\overline{F1}=0.92$ for all augmentations
combined) in comparison to training without augmentation ($\overline{F1}=0.98$). We interpret this decrease to reflect the
regularizing effect of data augmentation that can generally be observed in a network with fixed model capacity when
fitting data of increasing variability. Interestingly, this decrease was most pronounced for the pre-REM class (see
figure~\ref{fig:f1vis}(c,f)), in contrast to the artifact class that could be fitted perfectly by
the network on the training set. Mean F1 scores on the validation set show that none of the data augmentation algorithms
could improve on the F1 score achieved by the network trained only with the best regularization parameters.
Consequently, we evaluated this network on the test set where we found a mean F1 score of $\overline{F1}=0.78$.

Comparing the regularized with the un-regularized results indicates that
regularization induced increases of the F1 scores in the minority classes.  None of the
augmentation techniques were able to improve the F1 scores of the individual classes,
except time shift augmentation, which increased the validation F1 score of pre-REM from $0.50$ to $0.52$.

The test set consisted of one mouse each from both dietary groups.
We explored their individual F1 scores and
did not find any deviations exceeding 2~\%.  In particular, the
pre-REM F1 score for the typical mouse was $0.48$; for the altered-diet mouse we
found $0.48$.

\subsection*{Reduction of the network to the standard stages}

Next, we studied the classification performance of our optimized network
architecture in distinguishing between the three standard sleep stages Wake, REM,
and NREM (see table~\ref{tab:f1scores3stages}). With neither regularization nor
augmentation, we again observed our network to accurately approximate the
training data; on the validation data, we found a mean F1 score of $\overline{F1}=0.94$.  Using
the best regularization from the optimization on five stages --
$p_{dropout}=0.2$, and $\lambda=10^{-4}$ -- increased the mean F1 score on the validation set
to $\overline{F1}=0.96$.  We accepted this network and found on the test set, reduced to 3 stages, a mean F1 score
of $\overline{F1}=0.95$.  A detailed analysis revealed that improvements were made in the REM
class, the most under-represented class in the reduced datasets.
\begin{table*}[h]
  \begin{center}
    \begin{tabular}{c|c|c|c|c|c}
	    training configuration & dataset & F1 score & F1 score & F1 score & F1 score\\
				   & & (mean) & Wake & REM & NREM\\\hline
	    \multirow{2}{*}{\shortstack{no regularization,\\ no data augmentation}} & train & 0.99 & 0.99 & 1.00 & 0.99\\
				   & valid & 0.94 & 0.98 & 0.90 & 0.96\\
				   \hline
	    \multirow{2}{*}{\shortstack{best regularization\\ no data augmentation,\\ \textit{Best Reduced Network}}} & train & 0.99 & 0.99 & 1.00 & 0.99\\
				   & valid & 0.96 & 0.98 & 0.92 & 0.97\\
				   & \textbf{test} & \textbf{0.95} & \textbf{0.98} & \textbf{0.92} & \textbf{0.96}\\\hline
    \end{tabular}
    \caption{F1 scores for the training, validation, and test set for predicting the three main stages Wake, REM, and NREM sleep at various configurations of the hyperparameters. The network was trained on the rebalanced training set \#2 (see table~\ref{tab:datasets}). Only the best configuration (best reduced network) was evaluated on the test set.}
    \label{tab:f1scores3stages}
  \end{center}
\end{table*}
\begin{figure*}
    \centering
    \includegraphics[width=0.7\linewidth]{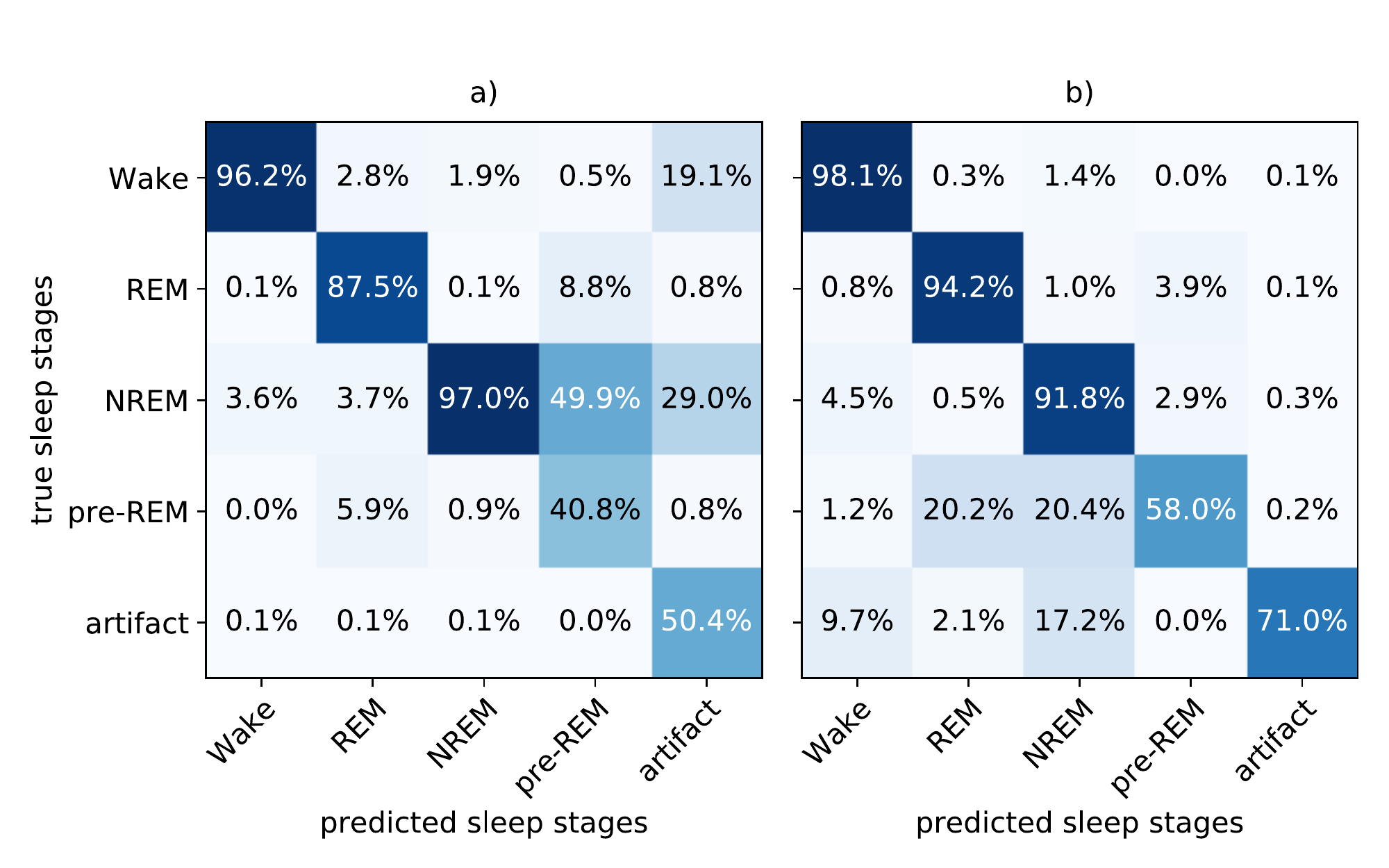}
    \caption{Confusion matrices of the best network evaluated on the test set (see table~\ref{tab:f1ScoresDataAug}): (a) Confusion matrix averaged by the predicted label frequency; diagonal elements correspond to precision.  (b) Confusion matrix averaged by true label frequency; diagonal elements correspond to recall. A prediction of pre-REM is about 50\% skewed towards NREM being the true stage, whereas a true pre-REM stage has equal probability to be mistaken for REM and NREM by the network.}
    \label{fig:confusion_matrices}
\end{figure*}

These classification performances are comparable with state-of-the-art results obtained in other studies\cite{Miladinovic2019,Barger2019,Svetnik2020} and demonstrate that the proposed network architecture is suitable for sleep scoring purposes.

\subsection*{Detailed analysis of the best network}

Confusion matrices (see section~\ref{sec:evaluation_measures}) obtained for the network trained without data augmentation (see figure~\ref{fig:confusion_matrices}) show that more than 98\% of the Wake segments, more than 94\% of the REM segments, and close to 92\% of the NREM segments in the test set were classified correctly (panel b). The network was also able to score 71\% of the artifact segments correctly. The pre-REM stage was scored in 58\% of the segments correctly but was confused in about 20\% of cases with REM or in 20\% of cases with NREM sleep.

The confusion matrices also show that more than 96\% of the segments in the test set predicted as Wake,
more than 87\% of the segments predicted as REM, and 97\% of the segments predicted as NREM had been assigned the
predicted stage by the human expert (panel a). Furthermore, more than 50\% of the segments that were predicted as
artifact by our network had been assigned artifact as the true stage. Predictions of pre-REM had pre-REM as the true
stage in nearly 41\% of cases, while in about 50\% of cases, predictions had NREM and about 9\% had REM as true stage.
\begin{figure*}
 \centering
 \includegraphics[width=0.7\linewidth]{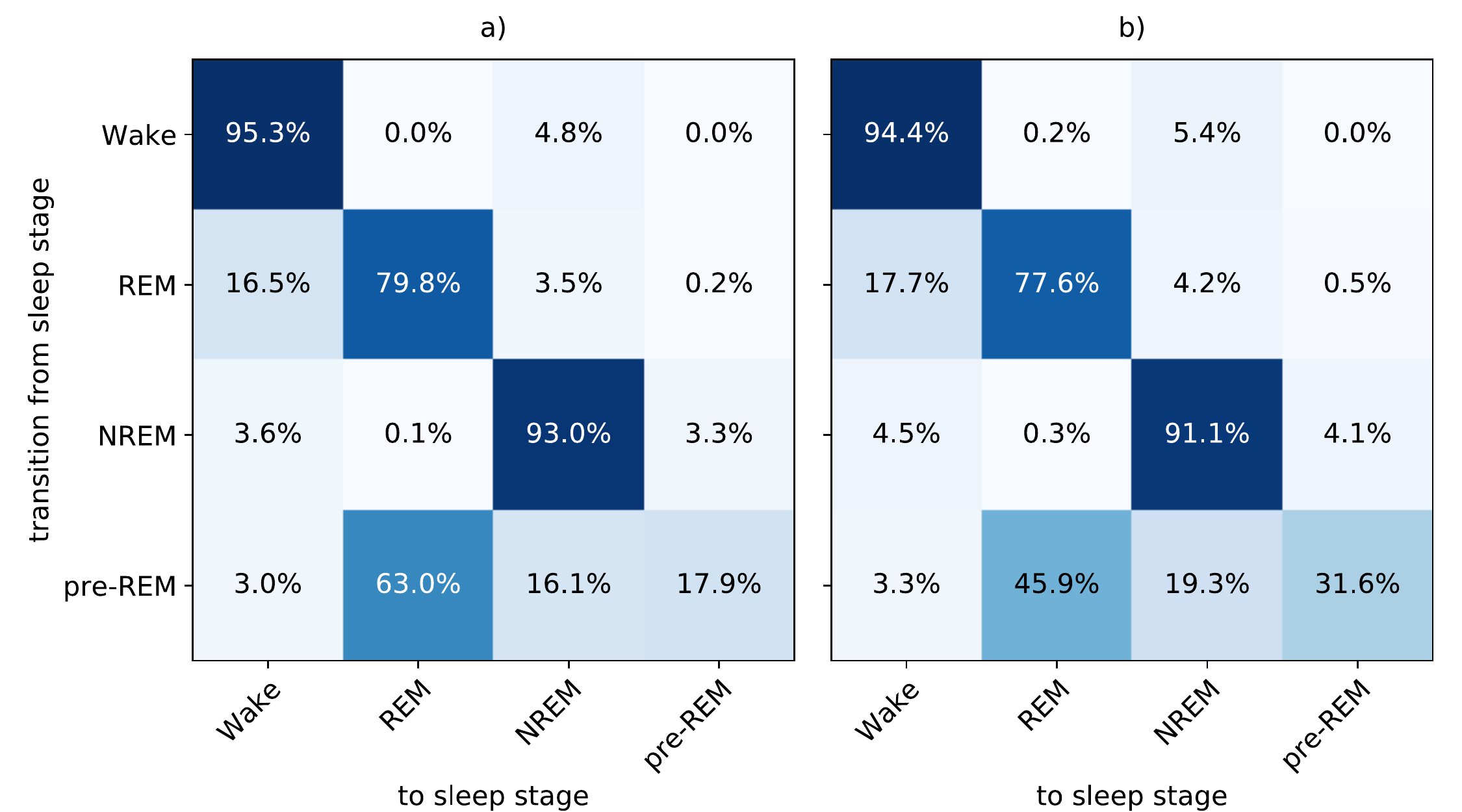}
 \caption{Transition probabilities from one stage (row) to another stage (column) obtained from the validation set with scores provided by our scoring expert (SW) (a) and obtained with scores predicted by the best network (b). The network parameters were trained on the rebalanced training set without augmentation.}
 \label{fig:tm}
\end{figure*}

To investigate whether the network was able to capture basic properties of sleep dynamics, we determined Markov transition matrices (see section~\ref{sec:evaluation_measures}) on the test set based on scores provided by our expert (figure~\ref{fig:tm}a) and on scores as predicted by our network (figure~\ref{fig:tm}b). The probabilities of transitions between the different stages obtained from the network indeed reflect typical sleep cycles, where Wake is followed by NREM sleep (5.4\%), which is followed by either Wake (4.5\%), pre-REM (4.1\%), or more NREM sleep (91.1\%). We highlight that there was only a small transition probability from NREM directly to the REM stage as determined by our network (0.3\%), which was close to the transition probability obtained by our expert (0.1\%). NREM transitioned via pre-REM to REM stages before the cycle closed with the transition to either Wake (17.7\%) or NREM (4.2\%). While transition probabilities of our network closely follow those as obtained from the expert, we find notable differences for transitions into and out of the pre-REM stage. pre-REM stages as determined by our network tended to last longer (pre-REM to pre-REM transition probability of 31.6\%) than those determined by the scoring expert (17.9\%).

\section{Discussion}

We proposed a neural network architecture that is able to distinguish between the three main stages Wake, REM and NREM as well as the infrequent stages pre-REM and artifact. When trained on these five stages, we observed our network to classify the majority sleep stages (Wake, REM, NREM) with high precision and recall (see figure~\ref{fig:confusion_matrices}), while the infrequent class pre-REM obtained lower precision and recall and a lower F1 score ($F1=0.48$, see table~\ref{tab:f1ScoresDataAug}) on out-of-sample data. Transition probabilities between sleep stages predicted by the network were largely consistent with those determined by the human expert. However, pre-REM phases predicted by the network tended to last longer than those by the expert. When the network was trained to distinguish between the three main stages only, classification performance as quantified by the average F1 score on out-of-sample data ($\overline{F1}=0.95$, see table~\ref{tab:f1scores3stages}) was comparable to state-of-the-art classifiers\cite{Miladinovic2019,Barger2019}. L2 and Dropout regularization as well as data augmentation increased F1 scores of the minority classes (pre-REM, artifact), while class rebalancing did not. We observed, however, the latter to stabilize the training process.

The F1 score of pre-REM indicates that pre-REM as a transitional sleep stage is particularly challenging to score. This challenge is not confined to scoring sleep in mice but can also be encountered in other species such as rats or humans where transitional sleep stages can also be observed. In rats, systems for sleep scoring obtained consistently lower scoring accuracies for the transition sleep stage compared to the main sleep stages\cite{Gross2009} with F1 scores ranging from $0.3$ to $0.7$\cite{Neckelmann1994,Wei2019}. In humans, automated scoring systems achieved F1 scores between $0.1$ to $0.6$ for N1 sleep\cite{Chambon2018,Supratak2017}, a transitional stage between wakefulness and sleep. In mice, we cannot comparatively assess the scoring performance of our network for pre-REM since, to the best of our knowledge, no prior approaches exist at the time of this writing. We note, however, that the obtained F1 score for pre-REM in mice ($F1=0.48$) is in the middle of the aforementioned ranges for F1 scores of transitional stages.

We hypothesize that pre-REM sleep is difficult to delineate from NREM and REM sleep for experts and automated systems alike. Indeed, figure~\ref{fig:data} (bottom row) shows a transition from NREM over pre-REM to REM sleep, where the predicted class probability of pre-REM already increases in the third segment that was labeled as NREM sleep. We observed many transitions like this where the class probabilities of NREM, pre-REM, and REM slowly decreased or increased along many segments. If human scorers had more difficulties in identifying pre-REM compared to other sleep stages, we would expect an increased intra-rater variance (i.e., a lower probability of labeling the same segments as pre-REM when the dataset would be labeled several times by the same person). Moreover, we would also expect larger inter-rater variance for pre-REM than for other stages (i.e., a lower agreement between different human scorers on scoring pre-REM) as has been observed for the transition sleep stage in rats\cite{Gross2009}. Such uncertainties in sleep scores (also called \emph{label noise})
can limit the classification performance achievable by automated scoring systems trained on that data. Besides label noise, another hypothesis that may explain lower classification accuracies for pre-REM is class imbalance\cite{Haixiang2017}. However, since we did not observe increased F1 scores when rebalancing classes (see figure~\ref{fig:f1vis}d and \ref{fig:f1vis}e), we reject this hypothesis for the learning problem and dataset studied here.

One of the limitations of this study is the lack of separate annotations by several experts from which a group consensus for each sleep stage could have been derived. Such annotations would have also allowed to assess the extent of inter-rater variance for each class, thereby allowing to judge the network's performance in classifying pre-REM sleep compared to human scorers. We note, however, that the scoring performance of the network when restricted to Wake, REM, and NREM is comparable to values reported for inter-rater agreement in the literature\cite{Barger2019}. Another limitation is that we could not investigate whether achievable F1 scores change due to age, behavior, disorders, dietary constraints, or genetic variations of mice since the dataset was too small. We expect some variability in F1 scores as indicated by previous studies (e.g., for genetic variations\cite{Franken1998,Miladinovic2019}) and consider future studies into this direction as desirable.

We consider several strategies as promising to improve upon existing data-driven systems for rodent sleep scoring like the neural network presented in this paper. (i) Statistical approaches such as \emph{confident learning}\cite{Northcutt2021,Lipton2018,Huang2019} aim at identifying mislabeled training instances and have been demonstrated to successfully increase classification performance in tasks such as image recognition\cite{Northcutt2021} by removing or correcting wrong labels, thereby reducing label noise. This approach might prove particularly fruitful for infrequent stages such as pre-REM sleep as we expect such stages to be more affected by label noise.
(ii) Constraining the sequence of predicted sleep stages to transition probabilities as observed in the training data\cite{Miladinovic2019} may improve network predictions for infrequent stages such as pre-REM.
We note, however, that such an approach can introduce bias in the frequencies of sleep stages when data was obtained under different experimental conditions. The reliable detection of such changes in sleep stage frequencies is, however, the matter of inquiry in many sleep studies.
(iii) When label noise can be reduced or when datasets scored by a group of experts (and thus with known inter-rater variabilities) become available, neural network architectures that could model long temporal successions\cite{Vaswani2017,Wu2020} of sleep stages might be particularly promising candidates to achieve new state-of-the-art classification performances for sleep scoring tasks.
(iv) Finally, due to the lack of consensus in rodent sleep scoring\cite{Robert1999}, we consider the creation of a public dataset annotated by a committee of experts to be particularly helpful to advance the creation of systems to automate scoring of animal sleep. Such a dataset should include data from different labs (to assess cross-lab variability of system predictions). Labels by a group of experts would establish a gold standard and would allow for an assessment of inter-rater variability, even for pre-REM sleep. Such approaches have been successfully pursued in other areas of sleep research, for instance for the challenge of sleep spindle detection\cite{Lacourse2020}.

We are confident that deep-learning based systems, such as introduced here, will facilitate sleep studies with large amounts of data, including long-term studies and studies with thousands of subjects. The automation of the manual sleep scoring process, a cognitive repetitive task, will likely support the reproducibility of studies and allow researchers and lab personnel to focus on productive tasks.

\section*{Acknowledgments}

We are grateful to M. Reißel and M. Grajewski for providing us with computing resources.

\section*{Author contributions}

Y.R. conceived the mice experiments and collected the data; S.W. annotated the data; J.T.C.S. and S.B. conceived the deep learning experiments; N.G. conducted the deep learning experiments, analyzed the results, and created the figures; S.B. wrote the first draft of the manuscript; all authors reviewed the manuscript.

\section*{Competing interests}

The authors declare no competing interests.

\end{document}